\documentclass[leqno,epsf,12pt]{amsart}
\input epsf.sty
\usepackage{amssymb}
\usepackage{amsfonts,amsthm,amscd}
\def\p{\;\raisebox{-1.5mm}{\epsfysize=6mm\epsfbox{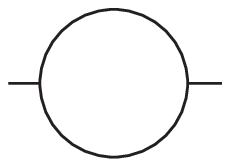}}\;}
\def\pdp{\;\raisebox{-1.5mm}{\epsfysize=6mm\epsfbox{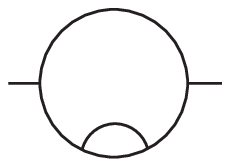}}\;}
\def\pdpdp{\;\raisebox{-1.5mm}{\epsfysize=6mm\epsfbox{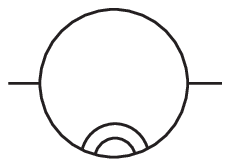}}\;}
\def\pdpdpdp{\;\raisebox{-1.5mm}{\epsfysize=6mm\epsfbox{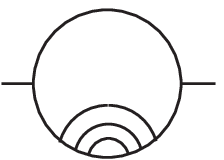}}\;}
\def\pddpp{\;\raisebox{-1.5mm}{\epsfysize=6mm\epsfbox{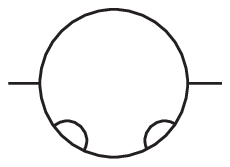}}\;}
\numberwithin{equation}{section}

\newtheorem{theorem}{Theorem}[section]
\newtheorem{corollary}[theorem]{Corollary}
\newtheorem{lemma}[theorem]{Lemma}
\newtheorem{proposition}[theorem]{Proposition}
\theoremstyle{definition}
\newtheorem{defn}[theorem]{Definition}
\theoremstyle{remark}
\newtheorem{remark}[theorem]{Remark}

\newcommand{\nc}{\newcommand}
\newcommand{\be}{\begin{equation}}
\newcommand{\ee}{\end{equation}}

\newcommand{\bc}{\begin{center}}
\newcommand{\ec}{\end{center}}
\nc{\bth}{\begin{theorem}} \nc{\bpr}{\begin{proposition}}
\nc{\epr}{\end{proposition}} \nc{\ble}{\begin{lemma}}
\nc{\ele}{\end{lemma}} \nc{\bco}{\begin{corollary}}
\nc{\eco}{\end{corollary}} \nc{\bre}{\begin{remark}}
\nc{\ere}{\end{remark}}
   \nc{\f}{\frac}
   \nc{\pa}{\partial}
   \nc{\na}{\nabla}
   \nc{\al}{\alpha}
   \nc{\bet}{\beta}
   \nc{\ga}{\gamma}
   \nc{\de}{\delta}
   \nc{\De}{\Delta}
   \nc{\Om}{\Omega}
   \nc{\om}{\omega}
   \nc{\tom}{\tilde\omega}
   \nc{\tq}{\tilde q}
   \nc{\ep}{\varepsilon}
   \nc{\tvp}{\tilde\varphi}
   \nc{\vp}{\varphi}
   \nc{\C}{\mathbb C}
   \nc{\R}{\mathbb R}

\begin{document}

\newpage\null\vskip-4em
  \noindent\scriptsize{}
  \scriptsize{
  \textsc{Center for Mathematical Physics BU-CMP/06-04} } 
\vskip 0.5 in

\normalsize

\title[Feynman diagrams and Lax pair equations]
{Feynman diagrams and Lax pair equations}
\date{\today}
\author{Gabriel B\u adi\c toiu}
\address{Department of Mathematics and Statistics, Boston
University, Boston, MA 02215. {\tt baditoiu@math.bu.edu}}
\author{Steven Rosenberg}
\address{Department of Mathematics and Statistics, Boston
University, Boston, MA 02215. {\tt sr@math.bu.edu}}

\begin{abstract}
We find a Lax pair equation corresponding to the Connes-Kreimer Birkhoff
factorization of the character group of the Hopf algebra of Feynman diagrams. 
In particular, we obtain a flow for the character given by Feynman rules,
and present a worked example.
\end{abstract}
\maketitle

\section{Introduction}
  In the theory of integrable systems, 
a solution to a Lax pair
  equation associated to a coadjoint orbit of a semisimple Lie group 
is given by a Birkhoff factorization on the corresponding loop group. 
By the work of Connes-Kreimer \cite{ck1}, 
there is a Birkhoff factorization of characters on the Hopf
  algebra of Feynman diagrams. 
In this paper, we reverse the usual procedure in
  integrable systems by
  producing 
 a Lax pair equation $\frac{d L}{dt}=[M,L]$ whose solution is
  given precisely by the Connes-Kreimer Birkhoff factorization (Theorem
  \ref{t:8.2}).
The main
  technical issue, that the Lie algebra of infinitesimal characters is not
  semisimple, is overcome by passing to the double Lie algebra with 
  the simplest possible Lie algebra structure.  In particular, 
  the Lax pair gives a flow for the character $\varphi$ given by
  Feynman rules in dimensional regularization.  It would be very
  interesting to know if this flow has physical significance.  

In \S\S1-4, we introduce a method to 
produce a Lax pair on any Lie algebra from equations of motion on the
double Lie algebra.
In \S5, we apply this method to the particular case of the Lie algebra of
infinitesimal characters of the Hopf algebra of Feynman diagrams,
and produce a
Lax pair equation whose Birkhoff factorization coincides with the
Connes-Kreimer factorization.  In \S6, 
we work out an explicit example of the theory 
on a finitely generated subalgebra of the Hopf algebra of Feynman diagrams.

It is natural to look for
  invariants of Lax pair equations
 by spectral curve techniques, and to linearize
  the flow on the Jacobian of the spectral curve. Unfortunately, in the worked
  example of \S6, the spectral curve is highly reducible, and the only
  invariants we find are trivial.  We hope to find examples with nontrivial
  invariants in the future.

\section{The double Lie algebra and its associated Lie Group}

There is a well known method to associate a Lax pair equation to a Casimir
element on the dual $\mathfrak g^*$
of a semisimple Lie algebra $\mathfrak g$
\cite{sts}. The semisimplicity is used
to produce an $\mathrm{Ad}$-invariant, symmetric,
non-degenerate bilinear form on $\mathfrak g$, allowing an identification of
$\mathfrak g$ with $\mathfrak g^*$.  For a general Lie algebra
 $\mathfrak g$,
there may be 
no $\mathrm{Ad}$-invariant, symmetric,
non-degenerate bilinear form on $\mathfrak g$. To produce a Lax pair,
we need to extend
 $\mathfrak g$ to a larger Lie algebra with such a 
bilinear form. We  do this by
constructing a Lie bialgebra structure on $\mathfrak g$ and extending
$\mathfrak g$ to $(\mathfrak g\oplus\mathfrak g^*,
[\cdot,\cdot ]_{\mathfrak g\oplus\mathfrak g^*})$, where
$[\cdot,\cdot ]_{\mathfrak g\oplus\mathfrak g^*}$ is the Lie
bracket induced by the Lie bialgebra. 

We recall the definition
of a Lie bialgebra structure (see e.g.~\cite{ksch}).
\begin{defn}\label{bialg}
  A Lie bialgebra is a Lie algebra $(\mathfrak g, [\cdot ,\cdot ])$
  with a linear map
  $\gamma:\mathfrak g\to\mathfrak g\otimes\mathfrak g$ such that
  \begin{itemize}
     \item[a)]
         $^t\gamma:\mathfrak g^*\otimes\mathfrak g^*\to\mathfrak g^*$ defines
         a Lie bracket on $\mathfrak g^*$,
     \item[b)]
         $\gamma$ is a $1$-cocycle of $\mathfrak g$, i.e.
         $$\mathrm{ad}^{(2)}_x(\gamma(y))-\mathrm{ad}^{(2)}_y(\gamma(x))-\gamma([x,y])=0,$$
         where $\mathrm{ad}^{(2)}_x:\mathfrak g^*\otimes\mathfrak
         g^*\to\mathfrak g^*\otimes\mathfrak g^*$ is given by
         $\mathrm{ad}^{(2)}_x(y\otimes z)=\mathrm{ad}_x(y)\otimes z+y\otimes
         \mathrm{ad}_x(z) 
= [x,y]\otimes z + y \otimes [x,z]$.
  \end{itemize}
\end{defn}

  A Lie bialgebra $(\mathfrak g,[\cdot,\cdot ],\gamma)$ induces an Lie
  algebra structure on the {\it double Lie algebra}
$\mathfrak g\oplus\mathfrak g^*$ by
  $$[X,Y]_{\mathfrak g\oplus\mathfrak g^*}=[X,Y],$$
  $$[X^*,Y^*]_{\mathfrak g\oplus\mathfrak g^*}= {}^t\gamma(X\otimes Y),$$
  $$[X,Y^*]=\mathrm{ad}^*_X(Y^*),$$
  for  $X$, $Y\in\mathfrak g$ and $X^*$, $Y^*\in\mathfrak g^*$,
  where $\mathrm{ad}^*$ is the coadjoint representation given by
  $\mathrm{ad}^*_X(Y^*)(Z)=-Y^*(\mathrm{ad}_X(Z))$ for
  $Z\in\mathfrak g$.

Since it is difficult to construct explicitly the Lie group associated to
the Lie algebra $\mathfrak g\oplus\mathfrak g^*$, we will
choose the trivial Lie bialgebra given by the cocyle $
\gamma=0$ and
denote by $\delta=\mathfrak g\oplus \mathfrak g^*$ the associated Lie
algebra. Let $\{Y_i,
i= 1,\ldots ,l\}$
be a basis of $\mathfrak g$, with dual basis
 $\{Y^*_i\}$.
The Lie bracket $[ \cdot,\cdot ]_\delta$  on $\delta$ is given by
 $$[Y_i,Y_j]_\delta=[Y_i,Y_j],\ [Y_i^*,Y_j^*]_\delta=0,\
   [Y_i,Y_j^*]_\delta=-\sum_k c^j_{ik}Y^*_k,$$
where the $c^j_{ik}$ are the structure constants: 
$[Y_i,Y_j]=\sum_kc^k_{ij}Y_k$.

The main point of this construction is that the natural pairing $\langle\cdot,
\cdot\rangle:\delta\otimes\delta\to\C$ given by $\langle Y_i,
Y_j^*\rangle=\delta_{ij}$ is an $\mathrm{Ad}$-invariant symmetric
non-degenerate bilinear form on the Lie algebra $\delta$. 

The Lie
group naturally associated to $\delta$ is given by the following
proposition.

\bpr\label{prop22}
   Let $\theta:g\times \mathfrak g^*\to \mathfrak g^*$ be
   the coadjoint representation
   $\theta(g,X)=\mathrm{Ad}^*_{g}(g)(X)$. Then the Lie algebra of the
   semi-direct product $\tilde G = 
g\rtimes_\theta \mathfrak g^*$ is the double
   Lie algebra $\delta$.
\epr

\begin{proof}
  The Lie group law on  the semi-direct product $\tilde
  G$
is given by
  $$(g,h)\cdot (g',h')=(gg',h+\theta_g(h')).$$
  Let $\tilde{\mathfrak g}$ be the Lie algebra of $\tilde G$. Then the
  bracket on $\tilde{\mathfrak g}$  is given by
  $$[X,Y^*]_{\tilde{ \mathfrak g}}
=d\theta(X,Y^*), \ \ [X,Y]_{\tilde {\mathfrak 
g}}=[X,Y],\ \  [X^*,Y^*]_{\tilde {\mathfrak g}}=0,$$
   for left-invariant vector fields $X$,
  $Y$ of $g$ and  $X^*, Y^*\in\mathfrak g^*$. We have
  $d\theta(X,Y^*)=d\mathrm{Ad}_{g}(X,Y^*)=[X,Y^*]_\delta$ since
  $d\mathrm{Ad}_{g}=\mathrm{ad}_{\mathfrak g}$.
\end{proof}

\section{The loop algebra of a Lie algebra}
Following \cite{adler}, we consider the loop algebra
$$L\delta=\{L(\lambda)=\sum\limits_{j=M}^N \lambda^jL_j \ |
                    \ M,N\in \mathbb Z, L_j\in\delta\}.$$
The natural Lie bracket on $L\delta $ is given by
$$\left[\sum \lambda^iL_i,\sum \lambda^j L_j'\right]=
\sum\limits_k \lambda^k\sum\limits_{i+j=k}[L_i,L_j'].$$
Set 
\begin{eqnarray*}L\delta _+ &=&
\{L(\lambda)= \sum\limits_{j=0}^N
\lambda^jL_j \ | \ N\in\mathbb Z^+\cup \{0\},  L_j\in\delta\}\\
L\delta _-&=&
\{L(\lambda)=\sum\limits_{j=-M}^{-1}   \lambda^jL_j \ | \ M\in\mathbb Z^+,
L_j\in\delta\}.
\end{eqnarray*}
Let $P_+:L\delta \to L\delta _+$ and $P_-:L\delta \to L\delta _-$ be
the natural projections and set $R=P_+-P_-$.

The natural pairing
$\langle\cdot,\cdot\rangle$ on $\delta$
extends to an $\mathrm{Ad}$-invariant, symmetric,
non-degenerate pairing on $L\delta $ by setting
   $$\left\langle\sum\limits_{i=M}^N \lambda^iL_i ,\sum\limits_{j=M'}^{N'}
     \lambda^jL'_j \right\rangle= 
     \sum\limits_{i+j=-1}\langle L_i,L_j'\rangle.
   $$

For our choice of basis $\{Y_i\}$ of $\mathfrak g$, we get an isomorphism
\begin{equation}\label{I}I: L\delta ^*\to L\delta \end{equation}
   with
      $$I\left(\sum L^j_iY_j\lambda^i\right)=\sum L^j_iY^*_j\lambda^{-1-i},
      $$

We will need the following lemmas.

\ble\cite{adler}
    We have the following natural identifications:
    $$L\delta _+=L\delta _-^* \ \mathrm{and\ } L\delta _-=L\delta _+^*.
    $$
\ele

\ble\cite[Lem.~4.1]{sts}\label{lem1}
    Let
    $\varphi$ be an $\mathrm{Ad}$-invariant polynomial on $\delta$. Then
      $$\varphi_{m,n}[L(\lambda)]=\mathrm{Res}_{\lambda=0}(\lambda^{-n}\varphi(\lambda^mL(\lambda))
      $$
    is an $\mathrm{Ad}$-invariant polynomial on $L\delta $ for $m, n\in\mathbb
    Z.$  
\ele

As a double Lie algebra,
$\delta$  has an   Ad-invariant polynomial, 
the quadratic polynomial
       $$\psi(Y)=\langle Y,Y\rangle       $$
associated to the natural pairing.
   Let $Y_{l+i}=Y^*_i$ for $i\in\{1,\ldots ,l\}$, 
so elements of $L\delta$ can be written
 $L(\lambda)=\sum\limits_{j=1}^{2l} \sum\limits_{i=-M}^N
    L_i^jY_j\lambda^i$.
Then the Ad-invariant polynomials
\begin{equation}\label{psimn1}
       \psi_{m,n}(L(\lambda))=\mathrm{Res}_{\lambda=0}(\lambda^{-n}
\psi(\lambda^mL(\lambda)),
\end{equation}
defined as in Lemma \ref{lem1} are given by
\begin{equation}\label{psimn2}
       \psi_{m,n}(L(\lambda))=2\sum
\limits_{j=1}^l\sum\limits_{i+k-n+2m=-1} L_i^jL_k^{j+l}.
\end{equation}
Note that
    powers of $\psi$ are also $\mathrm{Ad}$-invariant polynomials on
    $\delta $, so
\begin{equation}\label{psimnk}
      \psi^k_{m,n}(L(\lambda))=\mathrm{Res}_{\lambda=0}(\lambda^{-n}\psi^k(\lambda^mL(\lambda))
\end{equation}
    are  $\mathrm{Ad}$-invariant polynomials on $L\delta $.

It would be interesting to classify all Ad-invariant polynomials on $L\delta$
in general.

\section{The Lax pair equation}\label{s:7}
From \cite[Theorem 2.1]{sts}, the equation of motion induced by a
Casimir function $\varphi$ on the dual of a Lie algebra $\mathfrak h$ 
is given by
\begin{equation}\label{e:5.1}
    \frac{dL}{dt}=-\mathrm{ad}^*_{\mathfrak h}M\cdot L,
\end{equation}
where
$ L\in\mathfrak h^*, M=\frac{1}{2}R(d\varphi(L))\in\mathfrak h,
R=P_+-P_-$, where  $P_+$, $P_-$ are endomorphisms of
$\mathfrak h$ such that
     $$[X,Y]_R=[P_+X,P_+Y]-[P_-X,P_-Y]
     $$
is a Lie bracket on $\mathfrak h$.

Now we take $\mathfrak h=(L\delta)^*=L(\delta^*)$. Let  $P_\pm$ 
 be the projections of $L\delta^*$ onto $L\delta^*_\pm$.
After identifying $L\delta ^*=L\delta $
and $\mathrm{ad}^*=\mathrm{ad}$ via the map $I$  in (\ref{I}),
the equation of motion (\ref{e:5.1})
 can be written in Lax pair form
\begin{equation}\label{e:5.2}
    \frac{{d}L}{{d}t}=[L,M], 
\end{equation}
where $ M=\frac{1}{2}R(I(d\varphi(L(\lambda))))\in L\delta,$
and $\varphi$ is a Casimir function on $L\delta^* = L\delta$.

Finding a solution for (\ref{e:5.2}) reduces to the Riemann-Hilbert
(or Birkhoff)
factorization problem. The following theorem is a corollary of
  \cite[Theorem 2.2]{sts}.

\begin{theorem}\label{t:7.4}
   Let $\varphi$ be a Casimir function on $L\delta$ and set
   $X=I(d\varphi(L(\lambda)))\in L\delta$, for $L(\lambda) = L(0)(\lambda)
\in L\delta$. Let
   $g_{\pm}(t)$ be the smooth curves in $L({\rm Lie}(\delta))$
which solve the factorization
   problem
   $$\exp(-tX)=g_-(t)^{-1}g_+(t),$$
   with $g_{\pm}(0)=e$, and with $g_+(t) = g_+(t)(\lambda)$
 holomorphic in $\lambda\in \C$ and
$g_-(t)$ a polynomial in $1/\lambda$ with no constant term.
Let
   $M=\frac{1}{2}R(I(d\varphi(L(\lambda))))\in L\delta$.
   Then the integral curve $L(t)$ of the Lax pair equation
      $$\frac{d L}{d t}=[M, L]
      $$
   is given by
      $$L(t)=\mathrm{Ad}^*_{L\tilde G}g_{\pm}(t)\cdot L(0).
      $$
\end{theorem}

\vspace*{12pt}

This Lax pair equation projects to a Lax pair equation on the loop algebra of
  the original Lie
  algebra $\mathfrak g.$  
  Let $\pi_1$ be either the projection of  $\tilde G$ onto $G$ or its
  differential from 
$\delta$ onto $\mathfrak g$.
  This extends to a projection of $L\delta$ onto
 $L\mathfrak g$.  
  The projection of (\ref{e:5.2}) onto $L\mathfrak g$ is
\begin{equation}\label{e:pi1}
  \frac{d(\pi_1( L(t))}{dt}=[\pi_1(L),\pi_1(M)],
\end{equation}
since $\pi_1 = d\pi_1 $ commutes with the bracket.  
Thus the equations of motion (\ref{e:5.2}) induce a Lax pair equation on
$L\mathfrak g$, although this is not the equation of motion for a Casimir on
$\mathfrak g^*.$

\begin{theorem} The Lax pair equation of Theorem \ref{t:7.4} projects to a Lax
  pair equation on $L\mathfrak g.$ 
\end{theorem}


\vspace*{12pt}
When $\psi_{m,n}$ is the Casimir function on
$L\delta$ given by
    (\ref{psimn1}), $X$ can be written nicely in terms of $L(\delta)$.
\begin{proposition}\label{p:5.3}
    Let $X=I(d \psi_{m,n}(L(\lambda)))$.
    Then
\begin{equation}\label{e:5.3}
    X=2\lambda^{-n+2m} L(\lambda).
\end{equation}

\end{proposition}

\begin{proof}
Write $
      L(\lambda) =\sum\limits_{i,j}L_i^j\lambda^{i}Y_j.$
  By formula (\ref{psimn2}), we have
  \begin{equation}\label{e:partialpsimn}
    \frac{\partial \psi_{m,n}}{\partial L_p^t}= \left\{
    \begin{array}{cc}
    2L_{n-1-2m-p}^{t+l}, \ \ \ \text{if} \ \ t\leq l\\
    2L_{n-1-2m-p}^{t-l}, \ \ \ \text{if} \ \ t>l.\\
    \end{array}
    \right.
  \end{equation}
  Therefore
      \begin{eqnarray}
           \nonumber  X&=&I(d\psi_{m,n}(L(\lambda)))
            =\sum\limits_{p,t}\frac{\partial \psi_{m,n}}{\partial L_p^t}\lambda^{-1-p}Y^*_t\\
          \nonumber   &=&
            2\lambda^{-n+2m}\sum\limits_p(\sum\limits_{t=1}^{l}L^{t+l}_{n-1-2m-p}Y_{t+l}\lambda^{n-1-2m-p}
            +\sum\limits_{t=l+1}^{2l}L^{t-l}_{n-1-2m-p}Y_{t-l}\lambda^{n-1-2m-p})\\
           \nonumber  &=&2 \lambda^{-n+2m}L(\lambda).
       \end{eqnarray}
\end{proof}

\section{The main theorem for Hopf algebras}

Let $H=(H, 1, \mu,\De,\ep,S)$ be a Hopf algebra over $\C$ and let
$H^*=(H^*,1_*, \mu_*=\De^t,\De_*=\mu^t,\ep_*=1^t,S^*=S^t)$ be the
algebraic dual of $H$.

\begin{defn}
  The {\bf character group} 
$\mathrm{Char}(H)$ of a Hopf algebra $H$ is
   the group of ``group-like elements'' of $H^*$ (see
  \cite{kas}):
  $$G=\{\phi\in H^*\ | \ \phi\not=0,\ \Delta_*\phi=\phi\otimes\phi\}.$$
  The group law is given by the convolution product and the unit element
  is $1_*$:
    $$
      (\psi_1\star\psi_2)(h)=\langle  \psi_1\otimes\psi_2,\De h\rangle,
    $$
  $$\langle 1_*,h\rangle=\varepsilon(h),\ \ \mathrm{for\ } h\in H.$$
\end{defn}

\begin{defn}
  An {\bf infinitesimal character} of a Hopf algebra $H$ is a
  $\C$-linear map $Z:H\to\C$ satisfying
  $$\langle Z,hk\rangle =\langle Z, h\rangle \varepsilon(k)+\varepsilon(h)
  \langle Z,k\rangle.$$ The set of infinitesimal characters is denoted
  by $\partial\mathrm{Char}(H)$ and is endowed with a Lie algebra bracket:
  $$[Z,Z']=Z\star Z'-Z'\star Z,\ \  \mathrm{for\ }Z,\ Z'\in\partial Char(H), $$
  where
  $\langle Z\star Z',h\rangle=\langle Z\otimes Z',\Delta(h)\rangle$.
\end{defn}

Let $(H, 1=\emptyset, \mu,\De, \ep, S)$ be the universal Hopf algebra of all
Feynman graphs. 
For computations later, we also consider a Hopf subalgebra $H_0$ 
generated by a
finite number of Feynman graphs $A_1$, $A_2, \ldots, A_{l-1},
A_l=\emptyset$, and let $H_1$ be the Hopf subalgebra generated by
$A_1,\ldots, A_{l-1}$.   Let $G$ be the Lie group of characters of 
$H$, and let $G_0$ be the Lie group of characters of $H_0$.
The Lie algebra of infinitesimal characters $\mathfrak
g, \mathfrak g_0$ of $H, H_0$ are precisely the Lie algebras of $G, G_0$,
respectively.  

Let $Z_0$ be the infinitesimal character of $H$ given by
$Z_0(h)=\varepsilon(h)$.  For any generator of $H$, viz.~$T\in
\{A_1,\ldots A_{l-1},A_l=\emptyset\}$, let $Z_T$ be the
infinitesimal character given by $Z_T(T')=\delta_{T,T'}$. Notice
that $Z_0=Z_\emptyset$. The Lie algebra $\mathfrak g_0$ is generated
by $Z_0,\ Z_{A_1},\ldots,\ Z_{A_{l-1}}$. Let $\mathfrak g_1$ be the
Lie subalgebra of $\mathfrak g_0$ generated by $Z_{A_1},\
Z_{A_2},\ldots Z_{A_{l-1}}$. Let $G_1$ be the Lie subgroup of $G_0$
corresponding to $\mathfrak g_1$. Set $Y_i = Z_{A_i}$ for $i \in\{1,\ldots,
l-1\}$ and set $Y_l = Z_0.$  We also let $\mathfrak g_1$ denote the Lie
subalgebra of $\mathfrak g$ with basis $Z_T, T\neq \emptyset$, and let $G_1$
also denote the corresponding subgroup of $G$. 

\begin{remark}  We have
 $\mathfrak g = \mathfrak g_1 \oplus\langle
Z_\emptyset\rangle.$  At the Lie group level, 
 $G_0$ is the semi-direct product
   $G_1\rtimes\C$ given by
  $$(g,t)\cdot (g',t')=(g\cdot\theta_t( g'),t+t'),$$ where
  $\theta_t(g)(\Gamma)=e^{t\#(\Gamma)}g(\Gamma)$ for $\Gamma\in
  H_1$, and $\#(\Gamma)$ is the number of independent loops of
  $\Gamma$.  
\end{remark}



Recall that for any Lie group $K$, an element $L(\lambda)\in
LK$ has a Birkhoff decomposition if $L(\lambda) = L(\lambda)
_-^{-1}L(\lambda)_+$ with $L(\lambda)_-^{-1}$ holomorphic in $\lambda^{-1} \in
\mathbb P -\{0\}$ and $L(\lambda)_+$ holomorphic in $\lambda \in\mathbb P -
\{\infty\}.$  In the next lemma, $\tilde G$ refers either to 
$ G_0\rtimes_\theta \mathfrak
g^*_0$ as in
Prop.~\ref{prop22} or to $G\rtimes_\theta \mathfrak g^*$.

\ble\label{l:6.1}
    Let $(g,\alpha)$ be an element in $L\tilde G$. If
    $(g,\alpha)=(g_-,\alpha_-)^{-1}(g_+,\alpha_+)$ then $g=g_-^{-1} g_+$
    and $\alpha=\mathrm{Ad}^*(g_-^{-1})(-\alpha_-+\alpha_+).$
\ele
\begin{proof}
    We recall that
$(g_1,\alpha_1)(g_2,\alpha_2)=
(g_1g_2,\alpha_1+\mathrm{Ad}^*(g_1)(\alpha_2)).$
    Notice that
$(g_-,\alpha_-)^{-1}
=(g_-^{-1},-\mathrm{Ad}^*(g_-^{-1})(\alpha_-)),$ so
    $(g_-,\alpha_-)^{-1}(g_+,\alpha_+)=(g_-^{-1}
    g_+,-\mathrm{Ad}^*(g_-^{-1})(\alpha_-)+\mathrm{Ad}^*(g_-^{-1})(\alpha_+)$.
\end{proof}

We prove the existence of a Birkhoff
decomposition for any element
 $(g,\alpha)\in L\tilde G$.


\begin{theorem}
Every   $(g,\alpha)\in L\tilde G $ has a Birkhoff
decomposition 
  $(g,\alpha)=(g_-,\alpha_-)^{-1}(g_+,\alpha_+)$ with
  $(g_+,\alpha_+)$  a polynomial in $\lambda$ and
  $(g_-,\alpha_-)$  a polynomial in $\lambda^{-1}$
  without constant term.
\end{theorem}
\begin{proof}
    Let
       $g=g_-^{-1}g_+$
    be the Birkhoff decomposition of $g$ in $LG_0$ given in \cite{egk}.
    Let $\alpha_+=P_+(\mathrm{Ad}^*(g_-)(\alpha))$
    and
       $\alpha_-=-P_-(\mathrm{Ad}^*(g_-)(\alpha))$.
    Then, by Lemma \ref{l:6.1},
    $(g,\alpha)=(g_-,\alpha_-)^{-1}(g_+,\alpha_+)$.
\end{proof}

In \cite{ck1}, Connes and Kreimer give a Birkhoff decomposition for the
character group of the Feynman graph Hopf algebra, and in particular for 
the normalized
loop character $\bar\varphi(\lambda,q)$ of dimensional
regularization.
Here
    $$\bar\varphi(\lambda,q) = \f {\varphi(\lambda)}{q^2},
    $$
where $\varphi(\lambda, q)$ is the usual character given by dimensional
regularization and Feynman rules.
We consider the algebra of formal power series
    $$\Omega\delta=\{L(\lambda)=\sum\limits_{j=-\infty}^\infty \lambda^jL_j \ |
                    \ L_j\in\delta\}.
    $$
The natural Lie bracket on $\Omega\delta $ is 
     $$\left[\sum \lambda^iL_i,\sum \lambda^j L_j'\right]=
        \sum\limits_k \lambda^k\sum\limits_{i+j=k}[L_i,L_j'].
     $$
Set
    \begin{eqnarray*}\Omega\delta _+ &=& 
\{L(\lambda)= \sum\limits_{j=0}^\infty
       \lambda^jL_j \ |  \ L_j\in\delta\}\\
\Omega\delta _- &=& \{L(\lambda)=\sum\limits_{j=-\infty}^{-1}
      \lambda^jL_j \ |  \    L_j\in\delta\}.
\end{eqnarray*}



Recall that $\pi_1$ denotes either the projection of the double Lie group
$\tilde G$ to its
first factor $G$, 
its differential, or its extention to the loop group and loop
algebra.  We denote the image of an element by adding a tilde,
e.g. $\pi_1(L(\lambda)) = \tilde L(\lambda).$

\begin{theorem}\label{t:8.2}
   Let $H$ be the Hopf algebra of
   Feynman graphs, and let
   $\bar\varphi(\lambda)\in\mathrm{Char}(H) = G$.
 Set
 $$L_0(\lambda)=\frac{1}{2}\lambda^{n-2m}\exp^{-1}(\bar\varphi(\lambda))
     \ \text{and}\ \ \
      X=I(d\psi_{m,n}(L_0(\lambda)),
     $$
     where $\psi_{m,n}$ is the Casimir function on $\Omega\delta$
     given by
     $$\psi_{m,n}(L(\lambda))=\mathrm{Res}_{\lambda=0}
       (\lambda^m\langle\lambda^n L(\lambda),\lambda^n L(\lambda)\rangle).
     $$
     Then $\exp(X)=\bar\varphi$ and the solution of
   \begin{equation}\label{8:1}
        \frac{dL}{dt}=[M,L], \ \ \ M=\frac{1}{2}
        R(I(d\psi_{m,n}(L(\lambda)))
   \end{equation}
   with the initial condition $L(0)=L_0$ is given by
   \begin{equation}\label{e:8.3}
        L(t)=\mathrm{Ad}^*_{L\tilde G}g_{\pm}(t)\cdot L_0,
   \end{equation}
   with $\exp(-tX)=g_-(t)^{-1}g_+(t)$. Applying $\pi_1$, we get a Lax pair
   equation
$$\frac{d\tilde L}{dt} = [\tilde M, \tilde L]$$
on $LG$, with solution 
$\tilde L(t)=\mathrm{Ad}^*_{L G}\tilde g_{\pm}(t)\cdot \tilde L_0.$
In particular,
     $$\bar\varphi = \exp(X)=\tilde g_-(-1)^{-1}\tilde g_+(-1)
     $$
   is the Connes-Kreimer factorization of $\bar\varphi$.
The same
   results hold for any finitely generated Hopf subalgebra $H_0$
of $H$.
\end{theorem}

\begin{proof}


By \cite[p.~32]{man}, the exponential map $\exp:\mathfrak g \to G$ is
bijective,
so $L_0(\lambda)$ exists.
The theorem then
 follows from Theorem \ref{t:7.4} and (\ref{e:pi1})
applied to the natural pairing on
$\delta$ as in (\ref{psimn1}) and the uniqueness of the Birkhoff
factorization.  
\end{proof}


\begin{remark}  We have to be careful passing from finite dimensional Lie
  algebras to the infinite dimensional Lie algebra $\mathfrak g$ 
of infinitesimal 
  characters on $H$, since the finite dimensional
   identification of a Lie algebra
$\mathfrak h$ with $\mathfrak h^*$  via a choice of basis
   of $\mathfrak h$ may not be valid.
However, we  just need
  the part of $\mathfrak g^*$ spanned by the dual basis to a fixed basis of
  $\mathfrak g$, so the finite dimensional theory extends.
\end{remark}

\section{A worked example}

In this section we discuss a specific example of the main theorem.

Let
\begin{eqnarray*}
    A= \p\ ,\ AA=\pdp\ ,\ AAA=\pdpdp\ ,\ AAAA=\pdpdpdp \ , \ B=\pddpp\ .
\end{eqnarray*}

Let $H_0$ be the Hopf subalgebra generated by $\emptyset$, $A$,
$AA$, $AAA$ , $AAAA$, $B$, and let $H_1$ be the Hopf subalgebra generated by
$A$,
$AA$, $AAA$ , $AAAA$, $B$. Let $G_0, G_1$ be the Lie groups of
characters of $H_0, H_1$, respectively. 
The Lie algebra of infinitesimal characters
$\mathfrak g_0$ of $H_0$ is the Lie algebra of $G_0$, and similarly for
$\mathfrak g_1.$  In particular, the Lie
algebra $\mathfrak g_1$ of $G_1$ is generated by $Z_A,\ Z_{AA},\
Z_{AAA},\ Z_{AAAA},\ Z_B$. We identify $G_1$ with $\C^5$ using the
normal coordinates defined in \cite{chr}. The group law
induced on $\C^5$ is given by the following lemma.

\ble
 Let $\oplus:\C^5\times\C^5\to\C^5$ be the group law on $\C^5$ given by
  $$
     (x_1,x_2,x_3,x_4,x_5) \oplus (y_1,y_2,y_3,y_4,y_5)=(x_1+y_1,x_2+y_2,
      x_3+y_3,x_4+y_4+x_1y_2-x_2y_1,x_5+y_5).
  $$
 Define $F:G_1\to (\C^5,\oplus)$ by
    $$
      F(\varphi)=(\varphi(A),
                  \varphi(AA)-\f 12\varphi(A)^2,
                  \varphi(AAA)-\varphi(A)\varphi(AA)+ \f 16 \varphi(A)^3,$$ $$
                   \varphi(B)-\varphi(A)\varphi(AA)+\f 16\varphi(A)^3,
             \varphi(AAAA)-\varphi(A)\varphi(AAA)- \f 12 \varphi(AA)^2
                  +\varphi(A)^2\varphi(AA)-\f 14 \varphi(A)^4
                 ).
    $$
 Then F is a group isomorphism.
\ele

\begin{proof}
   Let $(F_1,F_2,F_3,F_4,F_5)=F$. We have
      $$\De(A)=A\otimes 1+1\otimes A,
      $$
      $$\De(AA)=AA\otimes 1+1\otimes AA+A\otimes A,
      $$
      $$\De(AAA)=AAA\otimes 1+1\otimes AAA+AA\otimes A+A\otimes AA,
      $$
      $$\De(AAAA)=AAAA\otimes 1+1\otimes AAAA+AAA\otimes A+A\otimes
      AAA+AA\otimes AA,
      $$
      $$\De(B)=B\otimes 1+1\otimes B+2A\otimes AA.
      $$
Therefore
      $$(\phi_1\star\phi_2)(A)=\langle \phi_1\otimes\phi_2,\De(A)\rangle=
        \phi_1(A)+\phi_2(A),
      $$
\begin{eqnarray*}
   (\phi_1\star\phi_2)(AA)-\f 12(\phi_1\star\phi_2)(A)^2
   &=&\langle \phi_1\otimes\phi_2,\De(AA)\rangle-\f 12(\phi_1\star\phi_2)(A)^2\\
   &=&
      \phi_1(AA)+\phi_2(AA)+\phi_1(A)\phi_2(A)-\f 12(\phi_1(A)+\phi_2(A))^2 \\
   &=&(\phi_1(AA)-\f 12\phi_1(A)^2)+(\phi_2(AA)-\f 12\phi_2(A)^2),
\end{eqnarray*}
  which implies $F_k(\phi_1\star\phi_2)=F_k(\phi_1)+F_k(\phi_2)$ for
  $k\in\{1,2\}$. By direct computations we also get
    $$F_k(\phi_1\star\phi_2)=F_k(\phi_1)+F_k(\phi_2)\ \  \mathrm{for}\
    k\in\{3,5\}, 
    $$
    $$F_4(\phi_1\star\phi_2)=F_4(\phi_1)+F_4(\phi_2)
      +F_1(\phi_1)F_2(\phi_2)-F_2(\phi_1)F_1(\phi_2).
    $$
\end{proof}

\medskip
Using this identification of $G_1$ with $(\C^5,\oplus)$, we can
identify $G_0$ with $(\C^6,\oplus )$, where
  $$
     (x_1,x_2,x_3,x_4,x_5,t)\oplus(y_1,y_2,y_3,y_4,y_5,t')=
  $$
  $$(x_1+e^ty_1,x_2+e^{2t}y_2,
      x_3+e^{3t}y_3,x_4+e^{3t}y_4+e^{2t}x_1y_2-e^{t}x_2y_1,x_5+e^{4t}y_5,t+t').
  $$
\noindent The following lemma gives a basis of the left invariant vector
fields on $G_0$ and the structure constants of $\mathfrak g_0$.

\ble \label{seven.two}
Let
  $$Y_1=e^t \left(\f\partial{\partial y_1}-y_2\f\partial{\partial y_4}\right), \ \ \ \
    Y_2=e^{2t} \left(\f\partial{\partial y_2}+y_1\f\partial{\partial y_4}\right),
  $$
  $$Y_3=e^{3t} \f\partial{\partial y_3}, \ \ \ \ Y_4=e^{3t} \f\partial{\partial y_4}, \ \ \ \
     Y_5=e^{4t}  \f\partial{\partial y_5}, \ \ \ \ Z_0=\f\partial{\partial t}.
  $$
    where $(y_1,y_2,y_3,y_4,y_5,t)$ are the standard coordinates on $\C^5$.
 Then $\{Y_1$, $Y_2$, $Y_3$, $Y_4$, $Y_5$, $Z_0\}$ is a basis of the left
 invariant vector fields on $G_0$. We have 
   $[Y_i,Y_j]=0$ for any $(i,j)\not=(1,2),(2,1)$, and
   $[Y_1,Y_2]=-[Y_2,Y_1]=2Y_4$, $[Z_0,Y_1]=Y_1$, $[Z_0,Y_2]=2Y_2$,
 $[Z_0,Y_3]=3Y_3$,  $[Z_0,Y_4]=3Y_4$ and $[Z_0,Y_5]=4Y_5$.
\ele

\begin{proof}
This follows from the easily computed formulas
    $$L_x\left(\f\partial{\partial y_1}\right)=e^t
      \left(\f\partial{\partial y_1}-x_2\f\partial{\partial y_4}\right),\ \
      L_x\left(\f\partial{\partial y_2}\right)=
      e^{2t}\left(\f\partial{\partial y_2}+x_1
      \f\partial{\partial y_4}\right),
    $$
    $$L_x\left(\f\partial{\partial y_3}\right)=e^{3t}\left(\f\partial{\partial y_3}\right),\ \
      L_x\left(\f\partial{\partial y_4}\right)=e^{3t}\left(\f\partial{\partial y_4}\right),\ \
      L_x\left(\f\partial{\partial y_5}\right)=
      e^{4t}\left(\f\partial{\partial y_5}\right),$$
$$\lefteqn{ L_x\left(\f\partial{\partial t'}\right)=\f\partial{\partial t'}.}\
      \ \ \ \ 
    $$
\end{proof}

\subsection{The exponential map and the adjoint and coadjoint representations}

\ble\label{l:1.5}
  The exponential $\exp:\mathfrak g_0\to G_0$ is
  given by
    $$\exp(a_1Y_1+a_2Y_2+a_3Y_3+a_4Y_4+a_5Y_5+a_6Z_0)=
    $$
    $$
      =\left\{
        \begin{array}{ll}
            (a_1,a_2,a_3,a_4,a_5,0), & \mbox{if } {a_6}=0\\
               \left(\frac{a_1(e^{a_6}-1)}{a_6}, \frac{a_2(e^{2a_6}-1)}{2a_6},
                      \frac{a_2(e^{3a_6}-1)}{3a_6}, \frac{a_4(e^{3a_6}-1)}{3a_6}
                     +\frac{a_1a_2}{2a_6^2}( \frac{e^{3a_6}-1}{3}-e^{2a_6t}+e^{a_6t}),
                      \frac{a_5(e^{4a_6}-1)}{4a_6},a_6
               \right), & \mbox{if } {a_6}\not=0,
        \end{array}
      \right.
   $$
  and $\exp$ is bijective.
\ele

\begin{proof}
  Let $Y\in\mathfrak g_0$ and  let $\gamma(t)$ be the $1$-parameter
  subgroup of $G_0$ generated by $Y$. Set
\begin{eqnarray*}Y &=& a_1Y_1+a_2Y_2+a_3Y_3+a_4Y_4+a_5Y_5+{a_6}Z_0,\\
\gamma(t) &=& (g_1(t),g_2(t),g_3(t),g_4(t),g_5(t),g_6(t)).
\end{eqnarray*}
  To find $\gamma(t)=\exp(tY)$, we solve the differential equation
  $$L_{\gamma(t)^{-1}}\dot\gamma(t)
        =a_1Y_1+a_2Y_2+a_3Y_3+a_4Y_4+a_5Y_5+{a_6}Z_0  $$
  with the initial condition $\gamma(0)=0$. First notice that
  $$\begin{array}{lll}
    L_{\gamma(t)^{-1}}\dot\gamma(t)&=& \dot g_1
    e^{-g_6}\f\partial{\partial h_1} +\dot g_2
    e^{-2g_6}\f\partial{\partial h_2} +\dot g_3
    e^{-3g_6}\f\partial{\partial h_3} +e^{-3g_6}(\dot g_1g_2-\dot g_2g_1
    +\dot g_4) \f\partial {\partial h_4} \\
   &&+\dot g_5e^{-4g_6}\f\partial{\partial h_5} +
   \dot g_6\f\partial{\partial h_6}.
     \end{array}
  $$
 \noindent Then $a_k=\dot g_ke^{-kg_6}$ for $k\in\{1,2,3\}$,
  $a_5=\dot g_5e^{-4g_6}$, $a_6=\dot g_6$ and $a_4=(\dot g_4+\dot
  g_1g_2-\dot g_2g_1)e^{-3g_6}$, with the initial conditions
  $g_1(0)=g_2(0)=g_3(0)=g_4(0)=g_5(0)=g_6(0)=0$.
  Therefore
  $g_6=a_6 t$, $g_k(t)=a_k(e^{ka_6t}-1)/(ka_6)$ for $k\in\{1,2,3\}$,
  $g_5(t)=a_5(e^{4a_6t}-1)/(4a_6)$
  and
  $$g_4=\frac{a_4(e^{3a_6t}-1)}{3a_6}-
    \int \frac{a_1e^{a_6t}a_2(e^{2a_6t}-1) } {2a_6}\ dt +\int
    \frac{a_2e^{2a_6t}a_1(e^{a_6t}-1)}{a_6}\ dt.
  $$

  $$g_4(t)=\frac{a_4(e^{3a_6t}-1)}{3a_6}+
   \frac{a_1a_2}{2a_6^2}\left(\frac{e^{3a_6t}-1}{3}-e^{2a_6t}+e^{a_6t}\right)
  $$
  If $a_6\not=0$ then
  $$\exp\left(\sum_{k=1}^5 a_kY_k + a_6Z_0\right)
     = \left(\frac{a_1(e^{a_6}-1)}{a_6}, \frac{a_2(e^{2a_6}-1)}{2a_6},
      \frac{a_2(e^{3a_6}-1)}{3a_6},\right.
  $$
  $$ \ \ \ \left. \frac{a_4(e^{3a_6}-1)}{3a_6}
     +\frac{a_1a_2}{2a_6^2}( \frac{e^{3a_6}-1}{3}-e^{2a_6t}+e^{a_6t}),
    \frac{a_5(e^{4a_6}-1)}{4a_6},a_6\right).
  $$
  If $a_6=0$ then
  $$\exp\left(\sum_{k=1}^5 a_kY_k\right)=(a_1,a_2,a_3,a_4,a_5,0).
  $$
\end{proof}

The adjoint and coadjoint representations of $G_0$ are given by the
following lemmas.

\ble\label{6.4}
  i) The adjoint representation $\mathrm{Ad}_{G_0}: G_0\to
  \mathrm{GL}({\mathfrak g_0})$ is given by
  $$
     \mathrm{Ad}_{ G_0}(g_1,\ldots, g_5, g_6)= \left(
     \begin{array}{cccccc}
     e^{g_6}      &0             &0 & 0     &  0       & -g_1   \\
     0            &e^{2g_6}      &0       & 0  & 0        &   -2g_2    \\
     0            &0             & e^{3g_6}& 0 & 0        & -3g_3\\
     -2g_2e^{g_6} &2g_1e^{2g_6}  &0          & e^{3g_6} & 0 & -g_1g_2-3g_4\\
     0            &0            &0          & 0         &  e^{4g_6}&-4g_5 \\
     0            & 0           & 0         & 0 &       0 & 1
     \end{array}
     \right)
  $$

   ii)  The coadjoint representation $\mathrm{Ad}^*_{G_0}: G_0\to
   \mathrm{GL}({\mathfrak g^*_0})$ is given by
   $$
         \mathrm{Ad}^*_{ G_0}(g_1,\ldots, g_5, g_6)=
         \left(
         \begin{array}{cccccc}
         e^{-g_6}      &0          &0 &  2g_2e^{-3g_6}   &  0 &0\\
         0            &e^{-2g_6}   &0 & -2g_1 e^{-3g_6}  & 0  & 0\\
         0            &0          & e^{-3g_6}& 0        & 0  &0\\
         0           &  0     &0              & e^{-3g_6}&   0 &0\\
         0            &0          &0   & 0           & e^{-4g_6} & 0\\
          e^{-g_6}g_1            &  2e^{-2g_6}g_2         &3e^{-3_6}g_3  &
          -e^{-3_6}g_1g_2+3e^{-3g_6}g_4 & 4e^{-4g_6}g_5  &1
         \end{array}
         \right)
   $$
\ele
\begin{proof}
     To show i), we straightforwardly 
compute $\mathrm{Ad}_{G_0}(g) =dC_g$ where
     $C_g(h)=ghg^{-1}$,
     %
 noting that $$ g^{-1} = (x_1,x_2,x_3,x_4,x_5,t)^{-1}=
     (-e^{-t}x_1,-e^{-2t}x_2,-e^{-3t}x_3,-e^{-3t}x_4,-e^{-4t}x_5,-t).$$
      $ii)$ then follows from  $i)$ and
     $\mathrm{Ad}^*(g)=(\mathrm{Ad}(g^{-1}))^t$.
\end{proof}

\ble
  1) For the basis $\{Y_1,\ldots,Y_6\}$ of Lemma \ref{seven.two},
$\mathrm{ad}:\mathfrak g_0\to
  \mathfrak{gl}(\mathfrak g_0)$ is given by

  $$
    \mathrm{ad}\left(\sum_{i=1}^6 c_iY_i\right)=
    \left(
    \begin{array}{cccccc}
      c_6     &0   &0  &0 & 0 &-c_1\\
      0      &2c_6 &0  &0 & 0 & -2c_2\\
      0      &0   &3c_6&0 &  0& -3c_3\\
     -2c_2& 2c_1 &0  &3c_6&  0  &-3c_4\\
      0      &0   &0   &0 &  4c_6 &-4c_5\\
     0 &0 &0 &0 &0 &0
    \end{array}
   \right)
  $$
 2) $\mathrm{ad}^*:\mathfrak g_0\to \mathfrak{gl}(\mathfrak g_0^*)$ is
 given by
 $$
   \mathrm{ad}^*\left(\sum_{i=1}^6 c_iY_i\right)= \left(
    \begin{array}{cccccc}
   -c_6   &0     &0      &2c_2    &0 &0\\
   0&  -2c_6    & 0      &-2c_1   &0 &0\\
   0&      0    &-3c_6   &  0     &0 &0\\
   0&      0     &0      &-3c_6   &0 &0\\
   0      &0     &0     &0 &  -4c_6  &0\\
   c_1 &   2c_2 &3c_3    & 3c_4 &4c_5  &0
   \end{array}
   \right)
 $$
\ele

Since $d$Ad $=$ ad, the proof follows by differentiating the 
formulas in Lemma \ref{6.4}.

\subsection{The double Lie algebra and its associated Lie group}

The
conditions a) and b) in Definition \ref{bialg}  of a Lie bialgebra  can be
written in a basis as a system of quadratic equations.
Solving this system explicitly, in our case via Mathematica, gives the
following proposition.

\bpr
     There are $43$ families of  bialgebra structures $\gamma$ 
on $\mathfrak g$.
\epr

In fact, the system of quadratic equations 
involves 90 variables.  Mathematica gives 
   1 solution with 82 linear relations (and so 8 degrees of freedom),
   7 solutions with 83 linear relations,
   16 solutions with 84 linear relations,
   13 solutions with 85 linear relations,
   5 solutions with 86 linear relations, and
   1 solution with 87 linear relations.

As it is difficult to construct explicitly a Lie group corresponding to the
Lie algebra $\mathfrak g\oplus\mathfrak g^*$ for an arbitrary choice of
$\gamma$, we take the simplest choice
$\gamma=0$ and let $\tilde G=G_0\rtimes\mathfrak g_0^*$ be the
corresponding Lie group of $\delta=\mathfrak g\oplus\mathfrak g^*$.

\begin{remark}
      The group law on $\tilde G$ is given by
      $$ ((g_1,g_2,g_3,g_4,g_5,g_6),(h_1,h_2,h_3,h_4,h_5,h_6))\cdot
         ((g'_1,g'_2,g'_3,g'_4,g'_5,g'_6),(h'_1,h'_2,h'_3,h'_4,h'_5,h'_6))=
      $$
      $$
        (g_1+e^{g_6}g'_1,
        g_2+e^{2g_6}g'_2, g_3+e^{3g_6}g'_3,
        g_4+e^{3g_6}g'_4+g^{2g_6}g_1g'_2-g^{g_6}g_2g'_1, g_5+e^{4g_6}g'_5,
        g_6+g'_6,$$ $$ h_1+e^{g_6}h'_1-2g_2e^{g_6}h'_4,
        h_2+e^{2g_6}h'_2+2g_2e^{2g_6}h'_4, h_3+e^{3g_6}h'_3,
        h_4+e^{3g_6}h'_4, h_5+e^{4g_6}h'_5,
      $$
      $$
      h_6-g_1h'_1-2g_2h'_2-3g_3h'_3-3g_4h'_4-4g_5h'_5+h'_6).
      $$
\end{remark}

\subsection{The adjoint representations  $\mathrm{ad}_\delta$ and
$\mathrm{Ad}_{\tilde G}$ }

Let $\mathrm{ad}_\delta:\delta\to 
{\mathfrak gl}(\delta)$ be the
adjoint representation of $\delta$. Computing $\mathrm{ad}_\delta$
explicitly, for example via Mathematica, we get
     $$\lefteqn{\mathrm{ad}_{\delta}\left(\sum\limits_{i=1}^{12}x_i Y_i\right)
              =}\ \      $$

$$ \left(
       \begin{array}{cccccccccccc}
          x_{6}& 0& 0& 0& 0& -x_{1}& 0& 0& 0& 0& 0& 0\\
          0& 2 x_{6}& 0& 0& 0& -2 x_{2}& 0&     0& 0& 0& 0& 0\\
          0& 0& 3 x_{6}& 0& 0& -3 x_{3}& 0& 0& 0& 0& 0& 0\\
          -2 x_{2}& 2 x_{1}& 0& 3 x_{6}& 0& -3 x_{4}& 0& 0& 0& 0& 0& 0\\
          0& 0& 0&     0& 4 x_{6}& -4 x_{5}& 0& 0& 0& 0& 0& 0\\
          0& 0& 0& 0& 0& 0& 0& 0& 0& 0& 0& 0\\
          0& -2 x_{10}& 0& 0& 0& x_{7}& -x_{6}& 0& 0& 2 x_{2}& 0& 0\\
          2 x_{10}& 0&     0& 0& 0& 2 x_{8}& 0& -2 x_{6}& 0& -2 x_{1}& 0&
0\\
          0& 0& 0& 0& 0& 3 x_{9}& 0&     0& -3 x_{6}& 0& 0& 0\\
          0& 0& 0& 0& 0& 3 x_{10}& 0& 0& 0& -3 x_{6}& 0&     0\\
          0& 0& 0& 0& 0& 4 x_{11}& 0& 0& 0& 0& -4 x_{6}& 0\\
          -x_{7}& -2 x_{8}& -3 x_{9}& -3 x_{10}& -4 x_{11}& 0& x_{1}& 2
x_{2}& 3
          x_{3}& 3 x_{4}& 4 x_{5}& 0
       \end{array}
   \right)
$$
where $Y_{6+t}=Y_t$ for $t\in\{1,\ldots,6\}$.

\vspace*{24pt}
\bco
      $\mathrm{Ker}(\mathrm{ad}_\delta)=\mathrm{Span}\{Y_{12}\}$.
\eco

\noindent The adjoint and coadjoint representations of ${\tilde G}$
are given in the following proposition.
 \bpr\label{p:3.2}
   $$
    \mathrm{Ad}_{\tilde G}(g_1,g_2,\ldots,g_{12})=
    \left(
     \begin{array}{cc}
      \mathrm{Ad}_{G_0}(g_1,g_2,\ldots,g_{6}) & 0\\
      H(g_1,g_2,\ldots,g_{12})& \mathrm{Ad}^*_{G_0}(g_1,g_2,\ldots,g_{6})
    \end{array}
   \right)
$$
$$
   \mathrm{Ad}^*_{\tilde G}(g_1,g_2,\ldots,g_{12})=
   \left(
    \begin{array}{cc}
      \mathrm{Ad}^*_{G_0}(g_1,g_2,\ldots,g_{6}) & H(g_1,g_2,\ldots,g_{12})\\
      0& \mathrm{Ad}_{G_0}(g_1,g_2,\ldots,g_{6})
    \end{array}
   \right)
$$
where
   $H(g_1,g_2,\ldots,g_{12})$ is a $6\times 6$ matrix given by
{\small
    $$
      \left(
       \begin{array}{cccccc}
      0 &       -2e^{2g_6}g_{10} & 0& 0 & 0& 4g_{10} g_2 + g_7\\
      2e^{g_6} g_{10}   & 0&      0&    0  &0& -2g_1 g_{10} + 2g_8\\
       0& 0 & 0& 0& 0 &3g_9\\
       0& 0 & 0& 0& 0 &3g_{10}\\
       0& 0 & 0& 0& 0 &4g_{11}\\
      6e^{g_6} g_{10} g_2 - e^{g_6} g_7  &-6e^{2{g_6}} g_1 g_{10} -
      2e^{2{g_6}}g_8 & -3e^{3 {g_6}} g_9& -3e^{3 {g_6}} g_{10} &-4e^{4
      {g_6}}g_{11} & z
      \end{array}
      \right)
    $$
}
and
    $$z=3g_1 g_{10} g_2 + 9 g_{10} g_4 +  16 g_{11} g_5 + g_1 g_7 + 4 g_2 g_8 + 9 g_3 g_9.
    $$
\epr

The proof is straightforward.

\subsection{Some $\mathrm{Ad}_{\tilde G}$-invariant polynomials on
$\delta$ and L($\delta$)}\label{s:5}

We note that $\mathrm{Tr}(\mathrm{ad}(a)^k), k\in \mathbb Z^+,$ 
are $\mathrm{Ad}$-invariant
polynomials on $\delta$. By Lemma \ref{lem1}, these induce
 $\mathrm{Ad}$-invariant polynomials on $L\delta$,
i.e.~Casimir
functions (constants of motions) on $L\delta $.
 Explicit computations give the following lemma.

\ble
      Let
      $\varphi:\delta\to\C$ be the map given by
      $\varphi(a)=\mathrm{Tr}(\mathrm{ad}(a)^k)$. 
For  odd
      positive integers $k$, 
$\varphi(a)=0$.
For even positive integers $k$,
      $\varphi(a)=C(a_6)^k$ 
for some constant $C = C_k$,
where      $a=\sum\limits_{i=1}^{12}a_iY_i$.
\ele

This gives the following corollary, which can also be checked directly.

\bco
  For any even positive  integer  $k$ and any constant $C$
   $$\varphi(a)=C(a_6)^k
   $$
  is an  $\mathrm{Ad}_{\tilde G}$-invariant polynomial on $\delta$.
\eco

\begin{proof}
      Let $\pi_6:\delta\to\C$ be the projection given by
\begin{equation}\label{e:pi6}
      \pi_6\left(\sum\limits_{j=1}^{12} a_iY_i\right)=a_6.
\end{equation}
        Since
      $$\pi_6(\mathrm{Ad}_{\tilde G}(\sum\limits_{j=1}^{12} a_iY_i))=a_6,$$
      we see that $\varphi(a)=C(a_6)^k$ is an $\mathrm{Ad}$-invariant
      polynomial on $\delta$.
\end{proof}

\subsubsection*{Example 1}
In the notation of (\ref{psimn1}), for integers $M$ and $N$ we get Casimir
elements on $L\delta$
    $$\varphi_{m,n}\left(\sum\limits_{i=M}^N\sum\limits_{j=1}^{12}
       L_i^jY_j\lambda^i\right)=\mathrm{Res}_{\lambda=0}(\lambda^{-n} C
       (\lambda^mL_6^i(\lambda^i))^{2k})
    $$
    $$=C\sum\limits_S
         \left(
          \begin{array}{cccc}
         &&k& \\
         i_M & i_{M+1} &\cdots &i_{N}\\
         \end{array}
         \right) (L^6_M)^{i_M}\cdots (L^6_N)^{i_{N}}
    $$
for all nonnegative
integers $m$, $n$, where
   $$S=\{(i_M,\ldots, i_N)\ | \
     i_M\geq 0,\ldots, i_N\geq 0,\ i_M+i_{M+1}+\ldots +i_{N}=k,
   $$
   $$Mi_M+(M+1)i_{M+1}+\ldots+Ni_N=-1+n-km\}.
   $$

\subsubsection*{Example 2}
The natural pairing $\langle\cdot,\cdot\rangle$ on $\delta $ induces
an $\mathrm{Ad}$-invariant polynomial
$$\psi(Y)=\langle Y,Y\rangle=2\sum\limits_{i=1}^6 a_ia_{i+6}$$  
for  $Y=\sum\limits_{i=1}^{12}a_iY_i$.
Then $\psi_{m,n}:L\delta\to\C$ given by (\ref{psimn1}) becomes
$$\psi_{m,n}(L(\lambda))=2\sum\limits_{j=1}^6\sum\limits_{i+k-n+2m=-1} L_i^jL_k^{j+6}.$$

\subsubsection*{Example 3}
(\ref{psimnk}) gives 
other $\mathrm{Ad}$-invariant polynomials on $L\delta$.
An explicit computation gives
$$\psi^k_{m,n}(L(\lambda))=\sum\limits_S
\left(
\begin{array}{c}
k\\
\{i_{a,b}\}_{a,b\in\{-M,\ldots,N\}}
\end{array}
\right) \prod\limits_{a,b\in\{-M,\ldots,N\}}
 (2\sum\limits_{j=1}^6 L_a^jL_b^{j+6})^{i_{a,b}},$$
where $$S=\{ \{i_{a,b}\}_{a,b\in\{-M,\ldots,N\}}\ |\
\sum\limits_{a,b\in\{-M,\ldots,N\}} i_{a,b}=k\ \ \mathrm{and}\
\sum\limits_{a,b\in\{-M,\ldots,N\}} i_{a,b}(a+b)-n=-1\}.
$$

\subsection{The Lax pair equation in local coordinates}

We can write the Lax pair equation
\begin{equation}\label{e:7.1}
        \frac{dL}{dt}=[M,L], \ \ \ M=\frac{1}{2}
        R(I(d\kappa(L(\lambda)))
\end{equation}
in local coordinates when the Casimir function $\kappa$ is
$\varphi_{m,n}$ or $\psi_{m,n}$ given above.

Note that
\begin{equation}\label{e:*}
 M=\frac 12 R(I(d\kappa(L(\lambda)))= \frac 12 \frac{\partial
(\kappa(L(\lambda))}{\partial L^t_p}Y_t \lambda^{-1-p} r(-1-p),
\end{equation}
 where
 $r(s)=1$ if $s\geq 0$ and  $r(s)=-1$ if $s< 0$.

\subsubsection*{The case
$\kappa =\varphi_{m,n}$}

\ble
      In local coordinates the Lax pair equation (\ref{e:7.1}) becomes
      $$\frac{d L^j_{i+p}}{dt}=0.$$
      for $0\leq j\leq 12$. Thus $L(t)=L(0)$ for any t.
\ele

\begin{proof}
          $$\frac{\partial \varphi_{mn}(\sum\limits_{i=M}^N\sum\limits_{j=1}^{12}
          L_i^jY_j\lambda^i)}{\partial L^6_p} =C\sum\limits_S
          \left(
           \begin{array}{cccc}
          &&k& \\
          i_M & i_{M+1} &\cdots &i_{N}\\
          \end{array}
          \right) (L^6_M)^{i_M}\cdots ((i_p)(L^6_p)^{i_p-1})\cdot
          (L^6_N)^{i_N},$$ where $S$ is some set of multi-indices. Then
          $$M=R(I(\frac{\partial \varphi_{mn}(\sum\limits_{i=M}^N\sum\limits_{j=1}^{12}
          L_i^jY_j\lambda^i)}{\partial L^6_p})Y_6\lambda^p)=r(-1-p)
          f(L(\lambda))Y_{12}.$$ Since $\mathrm{ad}(Y_{12})=0,$ we have
          $[L,M]=0$, and so $\frac{d L(t)}{d t}=0.$
\end{proof}

Thus the Lax pair equation is trivial in this case.

\subsubsection*{The case $\kappa =\psi_{m,n}$}

By Proposition \ref{p:5.3}, $M=R(X)=R(\lambda^{-n+2m}L(\lambda))$.

Let $q^k_{ij}$ be the structure constants of $\delta$ in the usual basis.

\begin{theorem}\label{t:7.3}
        In local coordinates the Lax pair equation (\ref{e:7.1}) becomes
        $$\frac{d L^k_{i+p-n+2m}}{dt}=r(-n+2m+p)
        \sum\limits_{j=1}^{12} \sum\limits_{t=1}^{12} L^j_iL^{t}_pq^k_{tj}$$
        for  $j\in\{1,\ldots,12\}$ and all $i,p$, where
         $r(s)=1$ if $s\geq 0$ and  $r(s)=-1$ if $s< 0$.
\end{theorem}

Note that the Lax pair equation $\pi_1(\dot L(t))=[\pi_1(L),\pi_1(M)]$
on $L \mathfrak g_0$ given by
(\ref{e:pi1}) has the local coordinate form
$$
     \frac{d L^k_{i+p-n+2m}}{dt}=r(-n+2m+p)
     \sum\limits_{j=1}^{6} \sum\limits_{t=1}^{6} L^j_iL^{t}_pc^k_{tj}
$$
for $j\in\{1,\ldots,6\}$ and all $i,p$.

\subsection{The Birkhoff factorization of  $\exp(-tX)$}

We compute  the factorization of $\exp(-tX)$ for the interesting case of
 $X=I(d
\psi_{m,n}(L(\lambda)))$, for $\pi_6(L(\lambda))=0$, where $\pi_6$
is the extension to $L\delta$ of $\pi_6$ in (\ref{e:pi6}).
 Then
$X=\lambda^{-n+2m}
L(\lambda)=\sum\limits_{j,i}(L_i^j\lambda^{i-n+2m})Y_j$, and
$$\exp(-tX)=\exp(\sum\limits_{j=1}^{12}(-\sum_i L^j_i\lambda^{i-n+2m}t)Y_j).$$
Let $z_j=-\sum\limits_i L^j_i\lambda^{i-n+2m}$ for 
$j\in\{1,\ldots, 12\}$.

Our assumption is that $z_6=0$, as the
 exponential of $ L(\lambda)$ has a simpler form in this case;
in fact this is the only case needed  for our main
theorem below. The exponential of $L\delta $ on $z_6=0$ is given by
$$\exp(tz_1,tz_2,tz_3,tz_4,tz_5,0,tz_7,tz_8,tz_9,tz_{10},tz_{11},tz_{12})=$$
$$\left(
tz_1,tz_2,tz_3,tz_4,tz_5,0, t^2 z_{10} z_2 + t z_7, -t^2 z_1 z_{10} + t z_8,
tz_9,tz_{10},tz_{11},\right.$$
$$\left.
t z_{12} - \frac{1}{3} t^3 z_1 z_{10} z_2 + \frac{3}{2} t^2 z_{10} z_4 +
2 t^2 z_{11} z_5 + \frac{1}{2} t^2 z_1 z_7 + t^2 z_2 z_8 +
\frac{3}{2} t^2 z_3 z_9\right).$$ 

By Lemma \ref{l:6.1}, the Birkhoff decomposition
$(g,\alpha)=(g_-,\alpha_-)^{-1}(g_+,\alpha)_+$, with $g\in G_0$ and
$\alpha\in \mathfrak g_0^*$ is given by $g=g_-^{-1}g_+$, and
$-\alpha_-+\alpha_+=\mathrm{Ad}^*(g_-)\alpha$.

Let $g_{j-}$ and $g_{j+}$,
$j\in\{1,\ldots, 12\}$, be the components of $g_-$ and $g_+$
respectively.
Therefore, for  $j\in\{1,2,3,4,5,9,10,11 \}$,
we have
$$g_{j+}=-\sum\limits_{i\geq n-2m} L^j_i\lambda^{i-n+2m}t$$ and
\begin{equation}\label{e:gj-}
    g_{j-}=-\sum\limits_{i< n-2m} L^j_i\lambda^{i-n+2m}t.
\end{equation}
Then
\begin{equation}\label{e:g-+}
   g_{i+}=P_+(tz_i), \ \    g_{i-}=-P_-(tz_i)
\end{equation}
for $i\in\{1,2,3,5\}$. We also get
\begin{equation}\label{e:g4+}
  g_{4+}=P_+(tz_4+t^2(P_-(z_1)P_+(z_2)-P_-(z_1)P_+(z_2))),
\end{equation}
\begin{equation}\label{e:g4-}
  g_{4-}=P_-(tz_4+t^2(P_-(z_1)P_+(z_2)-P_-(z_1)P_+(z_2))),
\end{equation}
\begin{equation}\label{e:g6-}
   g_{6-}=0,\ \ \ \ \ g_{6+}=0,
\end{equation}
\begin{equation}\label{e:g7-}
  g_{7-}=-P_-(2 t g_{2-} z_{10} + t^2 z_{10} z_2 + t z_7),
\end{equation}
\begin{equation}\label{e:g8-}
   g_{8-}=-P_-(-2 t g_{1-} z_{10} - t^2 z_1 z_{10} + t z_8),
\end{equation}
\begin{eqnarray}\label{e:g12-}
      g_{12-}&=&-P_-(-t g_{1-}g_{2-} 
z_{10} + 3 tg_{4-}z_{10}- 2 t^2 g_{2-}z_1 z_{10} +
      4t g_{5-} z_{11} \\
\nonumber      &&\qquad  + t z_{12} + t^2 g_{1-}z_{10} z_2 -
      \frac{1}{3}t^3z_1 z_{10}z_2
      +\frac{3}{2} t^2 z_{10} z_4 + 2 t^2 z_{11} z_5 + t g_{1-}z_7 \\
\nonumber     && \qquad +
      \frac{1}{2} t^2 z_1z_7 + 2t g_{2-} z_8 + t^2 z_2 z_8 + 3 t g_{3-} z_9 +
      \frac{3}{2} t^2 z_3 z_9 ),
\end{eqnarray}
$$g_{k+}=z_k+g_{k-},$$
for  $k\in\{7,8,12\}$. Then $g_-=(g_{1-},\ldots,g_{12-})$ and
$g_+=(g_{1+},\ldots,g_{12+})$ satisfy $g_-^{-1}g_+=\exp(-tX)$.

We  now assemble the final formulas needed to
compute the solution to the Lax pair equation
given by Theorem \ref{t:7.4}. Let
$\pi:\delta\to\mathfrak g$ and $\theta:\delta\to\mathfrak g^*$ be
the projections onto $\mathfrak g$ respectively $\mathfrak g^*$.
Then
\begin{equation}\label{e:adg}
       \pi(\mathrm{Ad}^*(g_{1-},\ldots, g_{12-})L(\lambda)))=
       \mathrm{Ad}^*_{G_0}(g_{1-},\ldots,
       g_{6-})\pi(L(\lambda)))+H(g_{1-},\ldots,g_{12-})\theta(L(\lambda))
\end{equation}
and
\begin{equation}\label{e:adg*}
\theta(\mathrm{Ad}^*(g_{1-},\ldots,g_{12-})L(\lambda)))=
    \mathrm{Ad}_{G_0}(g_{1-},\ldots, g_{6-})\theta(L(\lambda))),
\end{equation}
where $H$ is given by Proposition \ref{p:3.2}.
(It is fortunate  that
 $\mathrm{Ad}^*_{\tilde G}$ does not depend on $g_{12-}$, which by
(\ref{e:g12-}) is difficult
to compute explicitly.)

\subsection{The Feynman rules characters and the main result}

For any graph $\gamma$, the Feynman rules integral
 $\varphi(\lambda, q)(\gamma)$ can be
written  in term of $\Gamma$-functions. Some explicit formulas are as follows.

\ble
\begin{eqnarray*}
   \varphi(\lambda)(A) &=& \pi ^3 (q^2)^{1-\lambda}B_1(\lambda)\\
  \varphi(\lambda)(AA) &=& \pi^6(q^2)^{1-2\lambda}B_1(\lambda)B_2(\lambda)\\
\varphi(\lambda)(AAA) &=& 
\pi^9(q^2)^{1-3\lambda}B_1(\lambda)B_2(\lambda)B_3(\lambda)\\
\varphi(\lambda)(AAAA) &=&
  \pi^{12}(q^2)^{1-4\lambda}B_1(\lambda)B_2(\lambda)B_3(\lambda)B_4(\lambda)\\
\varphi(\lambda)(B) &=& \pi^9(q^2)^{1-3z} B_1(\lambda)^2B_2(\lambda)
\end{eqnarray*}
         where
           $B_j(\lambda)=\f{-1}{j\lambda(1-j\lambda)(2-j\lambda)(3-j\lambda)}
         \ , \ j\in\{1,2,3,4\}.$
\ele

\begin{theorem}
   Let $H_0$ be the Hopf subalgebra generated by $A$, $AA$, $AAA$, $AAAA$,
   $B$, $\emptyset$, and choose a character
   $\bar\varphi(\lambda)\in \Omega\tilde G$  with $\pi_6(\bar\varphi)=0$.
Set
     $$L_0(\lambda)=\frac{1}{2}\lambda^{n-2m}\exp^{-1}(\bar\varphi(\lambda)),
  \ \ \ 
      X=I(d\psi_{m,n}(L_0(\lambda)),
     $$
     where $\psi_{m,n}$ is the Casimir function on $\Omega\delta$
     given by
     $$\psi_{m,n}(L(\lambda)))=\mathrm{Res}_{\lambda=0}
       (\lambda^m\langle\lambda^n L(\lambda),\lambda^n L(\lambda)\rangle).
     $$
     Then $\exp(X)=\bar\varphi$ and the solution of
   \begin{equation}\label{e:8.1p}
        \frac{dL}{dt}=[M,L], \ \ \ M=\frac{1}{2}
        R(I(d\psi_{m,n}(L(\lambda)))
   \end{equation}
   with the initial condition $L(0)=L_0$ is given by
   \begin{equation}\label{e:8.3p}
        L(t)=\mathrm{Ad}^*g_{\pm}(t)\cdot L_0,
   \end{equation}
   with $\exp(-tX)=g_-(t)^{-1}g_+(t)$
   where $g_-$ and $g_+$ are given by  (\ref{e:gj-}), (\ref{e:g6-}),
   (\ref{e:g7-}), (\ref{e:g8-}), (\ref{e:g12-}),
 and where the $\mathfrak g_0$ and
   $\mathfrak g^*_0$ components of   $\mathrm{Ad}$ are given by 
(\ref{e:adg}) and (\ref{e:adg*}).

   In the particular case where $\bar\varphi$ is
   the normalized Feynman rule characters
   $\bar\varphi(\lambda)=\frac{\varphi(\lambda)}{q^2}$,
 $g_-(t)$ and $g_+(t)$ are given by (\ref{e:g-+}),
(\ref{e:g4+}) and    (\ref{e:g4-}),
and 
   the solution $L(t)$ of (\ref{e:8.1p}) is the flow of Feynman
   rules.
\end{theorem}


This follows from Theorem \ref{t:8.2}.


\begin{remark}
To any Lax equation with a spectral parameter, one can associate a
spectral curve and  study its algebro-geometric properties 
(see \cite{sts}).
In our case,  we consider the adjoint representation $\mathrm{ad}:\delta\to
{\mathfrak gl}(\delta)$ and the induced adjoint representation of the loop
algebra. The spectral curve is given by the characteristic equation
of $\mathrm{ad}(L\lambda)$
 $\Gamma_0=\{ (\lambda,\nu)\in\C-\{0\}\times\C\ | \
\det(\mathrm{ad}(L(\lambda)-\nu {\mathrm Id})=0\}$.

The theory of the spectral curve and its Jacobian usually assumes that the
spectral curve is irreducible.
Unfortunately, for all 43 bialgebra structures on $\delta$, on the associated
Lie algebra
 $\mathrm{ad}(\delta)$ all eigenvalues of the characteristic
equation 
are  zero, and the zero eigenspace is nine dimensional.  The spectral curve
itself 
is the union of degree one curves.
Thus each  irreducible
component has a trivial Jacobian, and the spectral curve theory
breaks down.
\end{remark}

\bibliographystyle{amsplain}
\bibliography{paper5}
\end{document}